\documentclass[aps,amsmath,amssymb,nofootinbib,superscriptaddress,sort&compress]{revtex4}
\usepackage{graphicx,hyperref}
\hypersetup{
  colorlinks   = true, 
  urlcolor     = blue, 
  linkcolor    = blue, 
  citecolor    = red   
}
\usepackage{hypernat} 

\begin{document}

\title{On the Second Law of Thermodynamics:\\ The Significance of Coarse-Graining and the Role of Decoherence}
\author{Mahdiyar Noorbala}
\affiliation{Department of Physics, University of Tehran, Tehran, Iran.  P.O.~Box 14395-547}
\affiliation{School of Astronomy, Institute for Research in Fundamental Sciences (IPM), Tehran, Iran.  P.O.~Box 19395-5531}

\begin{abstract}
We take up the question why the initial entropy in the universe was small, in the context of evolution of the entropy of a classical system.  We note that coarse-graining is a an important aspect of entropy evaluation which can reverse the direction of the increase in entropy, i.e., the direction of thermodynamic arrow of time.  Then we investigate the role of decoherence in the selection of coarse-graining and explain how to compute entropy for a decohered classical system.  Finally, we argue that the requirement of low initial entropy imposes constraints on the decoherence process.
\end{abstract}

\maketitle

\section{Introduction}

The second law of thermodynamics provides an arrow of time as the direction of the increase in entropy.  This implies that the initial entropy of our universe has been small and has led to discussions about the fine-tuning of the universe (see Ref.~\cite{Carroll} for a recent one).  In this paper we try to investigate the issue in the context of the evolution of entropy in the phase space.  In Section \ref{sec:cg} we emphasize the importance of coarse-graining and how it can change the direction of the arrow of time.  We show that the choice of coarse-graining corresponds to the choice of a coordinate grid.  Then in Section \ref{sec:dec} we explain how decoherence can fix this coordinate grid.  Then we come back to our original problem and see how the requirement of low initial entropy constrains the process of decoherence.  Finally, we summarize and conclude in \ref{sec:con}.

\section{The Significance of Coarse-Graining}\label{sec:cg}

\begin{figure}
\includegraphics[scale=1]{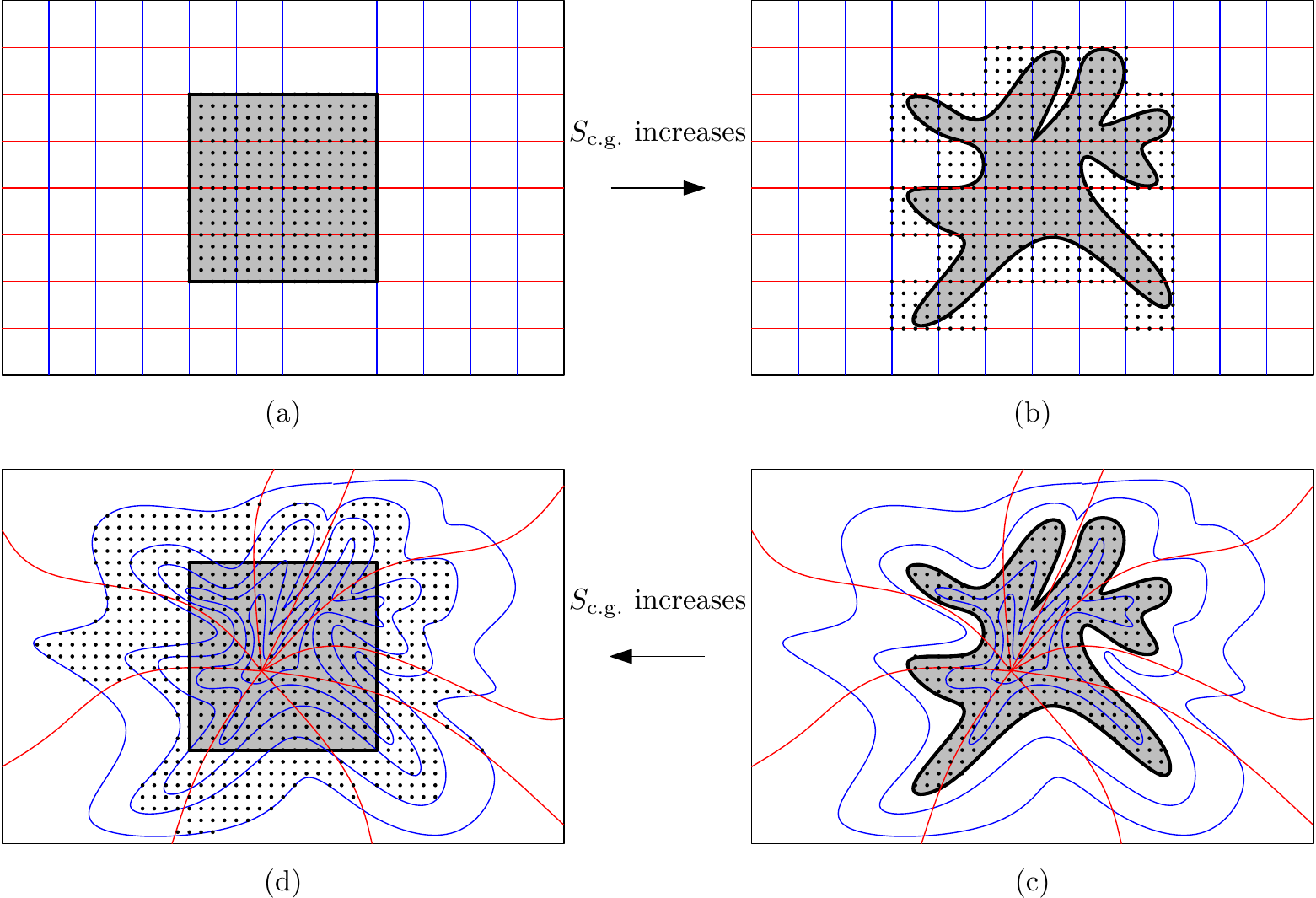}
\caption{Evolution of a distribution of points in phase space.  The distribution is uniform over gray regions and zero outside, so that entropy is directly proportional to the gray volume.  \textbf{(a)} The initial distribution in rectilinear coordinates.  There are 16 coordinate cells that are touched by the gray region.  \textbf{(b)} The final distribution in rectilinear coordinates.  Although the shape of the distribution is changed, its volume is the same.  The number of (dotted) coordinate cells that are touched by the gray region is now 27; coarse-grained entropy has increased.  \textbf{(c)} The same distribution as (b), but in a curvilinear coordinate system adapted to the shape of the distribution.  The number of (no longer box-shaped) coordinate cells that are touched by the gray region is 27 (the equality with the number in (b) is just a coincidence).  \textbf{(d)} The same distribution as (a), but in the curvilinear coordinates of (c).   The number of (dotted) coordinate cells that are touched by the gray region is now 40.  So the direction of the increase in coarse-grained entropy is opposite to that of the rectilinear coordinates.}
\label{fig:coarse-graining}
\end{figure}

According to the second law of thermodynamics, the entropy of a closed system increases with time.  One way to envision this in classical physics is to consider the phase space\footnote{Here and elsewhere in this paper, by ``phase space'' we mean the phase space of the entire system; so for example, a gas of $N$ atoms has a $6N$-dimensional phase space, not a 6-dimensional one.  We suppress the index $i$ on the canonical variables $q_i$ and $p_i$.} evolution of the system, where the Gibbs entropy of a distribution $f(q,p)$ on the phase space is given by
\begin{equation}
S = -\int f(q,p) \log f(q,p) dqdp.
\end{equation}
However, Liouville's theorem guarantees that any evolution governed by the classical equations of motion preserves $S$.\footnote{Strictly speaking, Liouville's theorem guarantees the conservation of the canonical volume element $dqdp$.  The conservation of $f$ in time then follows from the obvious fact that the total number of points $fdqdp$ in any volume element has to be conserved.}  Fig.~\ref{fig:coarse-graining} shows such a transition from an initial to a final distribution (a to b) with conserved entropy.  The element that turns $S$ to an increasing quantity is \textit{coarse-graining}.  To clarify, let us distinguish between two different notions that coarse-graining may refer to.  One is to introduce a few macroscopic variables to describe the system, rather than using all microscopic coordinates.  Another notion of coarse-graining, which is the one we use in this paper, is to introduce a degree of blurring and fuzziness on the shape of the evolving distribution, but keep the same number of variables.  Under this fuzziness, gray points that are close in Fig.~\ref{fig:coarse-graining}b but separated by white regions merge\footnote{At the heart of the coarse-graining procedure, we need to define a \textit{metric} (in the sense of distance, not inner product) and a \textit{resolution scale} on the phase space to judge when two points are `close' enough to be merged.  This metric space is further structure on top of the volume form which provides a \textit{measure} on the phase space.  We do not delve into these details.} to form a larger volume and hence give a larger entropy (since entropy is equivalent to phase space volume in this example).

One way to implement coarse-graining is to employ a coordinate system such that the characteristic size of the coordinate grid corresponds to the resolution scale of coarse-graining: the larger the grid size, the more coarse-grained will be the calculated volume/entropy.  This is done in Fig.~\ref{fig:coarse-graining}b (and trivially in \ref{fig:coarse-graining}a): the coarse-grained volume consists of all box-shaped coordinate cells that are fully or partially covered by the gray region.  It is then clear that ordinary and coarse-grained entropy are equal in Fig.~\ref{fig:coarse-graining}a, but coarse-grained entropy is larger than ordinary entropy in Fig.~\ref{fig:coarse-graining}b.  So the second law of thermodynamics holds, if we use the coarse-grained entropy and follow the arrow of time from a to b.

It is important to note how coarse-graining depends on the choice of coordinates.  In Figs.~\ref{fig:coarse-graining}c and \ref{fig:coarse-graining}d the coordinate grid is chosen to suit the `final' distribution.  Our coordinate cells (regions bounded by neighboring red `radial' and blue `circular' curves) are no longer box-shaped, and hence we have a different coarse-graining.  As a result, in Fig.~\ref{fig:coarse-graining}c we have a situation similar to Fig.~\ref{fig:coarse-graining}a, where ordinary and coarse-grained entropy are equal.  On the other hand, the coarse-grained entropy in Fig.~\ref{fig:coarse-graining}d is now larger than the ordinary entropy, in view of this new choice of coordinates.

We thus conclude that the direction of the increase in coarse-grained entropy (thermodynamic arrow of time) depends on the choice of coordinates used in coarse-graining.  In a sense, this is pretty natural.  Once a coordinate system is picked, the boundary of our distribution in the phase space can be described by an equation in terms of those coordinates.  The complexity of that equation---and hence of our description---is related to the degree of similarity between the shape of the boundary and the shape of the coordinate grid.  In other words, what seems very ordered in one coordinate grid (and requires very few bits to describe) may turn out to be very complicated in a different grid (and require many bits of information).

As a final remark, we note that it is always possible to choose a coordinate grid that assigns lower coarse-grained entropy to the initial (final) state compared to the final (initial) state.  It seems plausible that we can also find a grid which maintains a monotonic relation between coarse-grained entropy and time.  This is a harder task since it involves all times in between the initial and final moment, nevertheless we take it for granted without proof.

\section{The Role of Decoherence}\label{sec:dec}

As we saw in the preceding section, the choice of coordinates is fundamental in coarse-graining.  So which coordinate system should we choose?  Let us consider the case of a room filled with a gas of atoms, where our intuition can guide us.  The natural choice here is positions $q$ and momenta $p$.  If the atoms start out in a corner of the room, they begin to spread to other regions (both in space and phase space) but they won't occupy the entire phase space, because their phase space volume cannot increase.  However, if we choose this natural coordinate grid, then their coarse-grained volume/entropy does increase until a uniform distribution on the entire phase space is reached.  Why do we regard $q$ and $p$ as our natural coordinates?  

The underlying reason is that these are the quantities we can measure and probe the system with.  In other words, $q$ and $p$ are classical variables that correspond to pointer bases $\{|q\rangle\}$ and $\{|p\rangle\}$, respectively,  which are in turn selected via decoherence \cite{Zurek01}.  Of course, we usually measure only a few of these variables or particular combinations of them (like macroscopic variables such as the center of mass, volume, and pressure of the gas),\footnote{\label{fn} If, for a gas of $N$ atoms, we specify only 3 variables, then instead of a grid we will have foliations of the phase space with codimension-3 hypersurfaces blurred at the resolution scale.  A grid is the case of specifying $6N$ variables, yielding blurred points.} but we want to  see what we get if we measure everything we can---albeit with a given resolution.  One may legitimately employ canonical variables like $(q',p') = \left( \frac{q-p}{\sqrt2}, \frac{q+p}{\sqrt2} \right)$ to study the system;\footnote{But that doesn't mean a different pointer basis (they don't represent eigenvectors of, e.g., $Q\pm P$); and even if they did, $Q\pm P$ wouldn't be a classical variable selected by decoherence.} but using $(q',p')$ for coarse-graining doesn't have physical meaning, since our measurements only involve readouts of $q$ and $p$.  So we are left with $(q,p)$ as our coarse-graining coordinates.  

As our last takeaway from this example, we note that decoherence doesn't take place for all degrees of freedom; for example, although the positions of atoms become classical but the internal states of their electrons or quarks do not, unless specific measurements are performed to create appropriate interactions with the environment.  We exclude such non-classical variables from the phase space in our classical treatment of the system and incorporate them into the environment.  Again it is the decoherence process that determines which degrees of freedom become classical.  In other words, our probing options, i.e., available pointer bases, are determined by the nature of the environment and its constituent fragments as recorders of information about the system \cite{Zurek09}.  For the classical gas these are, for example, individual particles' positions, momenta, energies, and angular momenta, corresponding to pointer bases $\{|q\rangle\}$, $\{|p\rangle\}$, $\{|E\rangle\}$, and $\{|l,m_z\rangle\}$.  An important feature of the $\{|q\rangle\}$ and $\{|p\rangle\}$ bases is that they are eigenvectors of canonical operators $Q$ and $P$ (where all $Q$'s and all $P$'s commute among themselves, but $[Q_i,P_j]=i\hbar\delta_{ij}$) which generate the algebra of operators.  It seems that there is no other pair of pointer bases for this system that has this feature.  So we take as part of our proposal that the coarse-graining coordinates should correspond to pointer bases which arise from canonical operators.

We can now describe the situation in general.  To obtain the appropriate coordinates for coarse-graining, we go back to the time decoherence was first established and concentrate on those degrees of freedom which have become classical.  This provides a set of pointer bases, which we assume to come in the form of eigenvectors $\{|q\rangle\}$ and $\{|p\rangle\}$ of canonical operators $Q$ and $P$.  We then choose $(q,p)$ as our desired coordinates on phase space and compute the corresponding classical distribution from the quantum state of the system.  For example, for a density matrix $\rho$, we can use the Wigner distribution\footnote{The Wigner distribution can become negative (over regions of phase space with Planckian size), but if decoherence is established $f$ becomes positive \cite{Zurek03}.}
\begin{equation}
f(q,p) = \frac{1}{\pi\hbar} \int dq' \langle q+q' | \rho | q-q' \rangle e^{-2ipq'/\hbar}.
\end{equation}
At this stage, we compute the entropy of $f(q,p)$ coarse-grained in the $(q,p)$ grid.  For systems we observe this should be a small number, compared to the late time coarse-grained entropy, since we see the entropy increase in our world.  If the distribution $f(q,p)$ fits in the grid $(q,p)$ with few or no partially covered cells, then we have a situation with low initial entropy.  This seems to be the case for our universe, since it's apparently begun with little entropy.

A generic distribution $f(q,p)$ has wild behavior and is unlikely to be organized within the grid.  Thus the requirement of low initial entropy imposes constraints on the decoherence process through restrictions on the pointer bases it yields.  There are three elements in this process that may be constrained.  The first one is the quantum state $\rho$ before decoherence.  The second one is the system/environment splitting, especially since at times we incorporate into the environment some of the degrees of freedom that may appear to belong to the system (like the electron/quark states in the gas example above).  This is also the case if we want to study the initial entropy of the universe as a closed system, in which case a large number of degrees of freedom are included in the system but also a large number of them are reserved as environment (otherwise, we cannot think of this closed universe as a classical system).  Finally, the form of the Hamiltonian that describes the interaction between the system (or parts of it that we study) and the environment (and/or the rest of the system) is affected by these constraints.  In this context, therefore, the search for the reason behind the low entropy of the early universe is to be accompanied by investigations of the decoherence process.

\section{Conclusions}\label{sec:con}

We noticed the significance of coarse-graining in the definition of the entropy as it appears in the second law of thermodynamics.  We also pointed out the importance of the choice of the coordinate grid in coarse-graining, noting that the direction of increase in the coarse-grained entropy can be flipped at will by changing the grid.  Then we argued that decoherence is the underlying element the determines the grid and hence the arrow of time.  Finally, we explained how this leads to constraints on the decoherence process if we want to have a low initial entropy.

\medskip

\textbf{Note added.}  At the time of completion of this work, an independent paper by Rovelli with similar ideas appeared on arXiv \cite{Rovelli}.  While some of the basic ideas are shared in our works, there are differences as well.  The notion of coarse-graining employed in Ref.~\cite{Rovelli} is different from the one used here.  It is related to coarse-graining with fixing macroscopic variables, as discussed in the first paragraph of Section \ref{sec:cg} here.  Then the macroscopic observables that are to be used for coarse-graining are picked by interactions among subsystems.  Thus in Ref.~\cite{Rovelli}, interactions play the role that decoherence plays here.  As we discussed at the end of Section \ref{sec:dec}, decoherence plays this role through interactions as well as system/environment splitting and the quantum state.  So there is quite a similarity here.

\section{Acknowledgements}
I would like to thank Farbod Hassani, Behrad Taghavi, and Sarang Zeynizadeh for stimulating discussions.  I also acknowledge financial support from the research council of University of Tehran.


\begin{thebibliography}{9}

\bibitem{Carroll} 
	S.~M.~Carroll,
	``In What Sense Is the Early Universe Fine-Tuned?,''
	arXiv:1406.3057 [astro-ph.CO].

\bibitem{Zurek01}
	W.~H.~Zurek,
	``Decoherence, einselection, and the quantum origins of the classical,''
	Rev.\ Mod.\ Phys.\  {\bf 75}, 715 (2003)
	[arXiv:0105127 [quant-ph]].

\bibitem{Zurek09}
	W.~H.~Zurek,
	``Quantum Darwinism,''
	Nature Physics {\bf 5}, 181 (2009)
	[arXiv:0903.5082 [quant-ph]].
	  
\bibitem{Zurek03} 
	W.~H.~Zurek,
	``Decoherence and the transition from quantum to classical---Revisited,'' 
	arXiv:0306072 [quant-ph].

\bibitem{Rovelli} 
	C.~Rovelli,
	``Why do we remember the past and not the future? The `time oriented coarse graining' hypothesis,''
	arXiv:1407.3384 [hep-th].
	
\end{thebibliography}
\end{document}